\documentclass[slac_one]{revtex4}
\usepackage{graphicx}
\usepackage{fancyhdr}
\begin{document}

\title{
Emittance Measurement in MICE} 

%

\author{T. L. Hart and D. M. Kaplan}
\affiliation{Illinois Institute of Technology, Chicago, IL 60616, USA}
\begin{abstract}
 Muon ionization cooling provides the only practical solution to prepare high-brilliance beams necessary for a neutrino factory or muon collider. The Muon Ionization Cooling Experiment (MICE), under development at the Rutherford Appleton Laboratory, is installing the first set of particle detectors ever built to measure to 0.1\% the emittance of a 200\,MeV/$c$ or so muon beam in and out of a cooling cell, and thus measure the cooling cell's performance. Two identical ``emittometers" (a precise scintillating-fiber tracker in solenoidal magnetic field and a 50\,ps time-of-flight station) measure the six phase-space coordinates of each muon. Another TOF plane and two Cherenkov counters assure the purity of the incoming muon beam. A downstream electron/muon calorimeter eliminates contamination from decay electrons. 
\end{abstract}

\maketitle

\thispagestyle{fancy}


\section{INTRODUCTION} 
A muon collider~\cite{MC} may be the optimal way to study high-energy lepton-antilepton collisions, and  a neutrino factory~\cite{NF} based on a muon storage ring may be the ideal tool for the study of neutrino oscillation and the search for leptonic {\em CP} violation.  Ionization cooling~\cite{cooling,Neuffer}, while not yet experimentally demonstrated, is essential for high-luminosity muon colliders and has been shown by simulations and design studies to be an important factor for the performance and cost of a neutrino factory.  The international R\&D program towards neutrino factories and muon colliders has as a key goal  a first experimental demonstration of muon ionization cooling.  This is underway in the  international Muon Ionization Cooling Experiment (MICE)~\cite{MICE}, under construction at the UK's Rutherford Appleton Laboratory (RAL).

The main goals of MICE are to
\begin{itemize}
\item Design, engineer, and build a section of cooling channel with performance suitable for a neutrino factory;
\item Place the cooling channel in a muon beam and measure its performance in various operation modes and beam conditions, to test the limits and practicality of muon cooling.
\end{itemize}

\section{EXPERIMENT LAYOUT}

\begin{figure}
\includegraphics[width=6in]{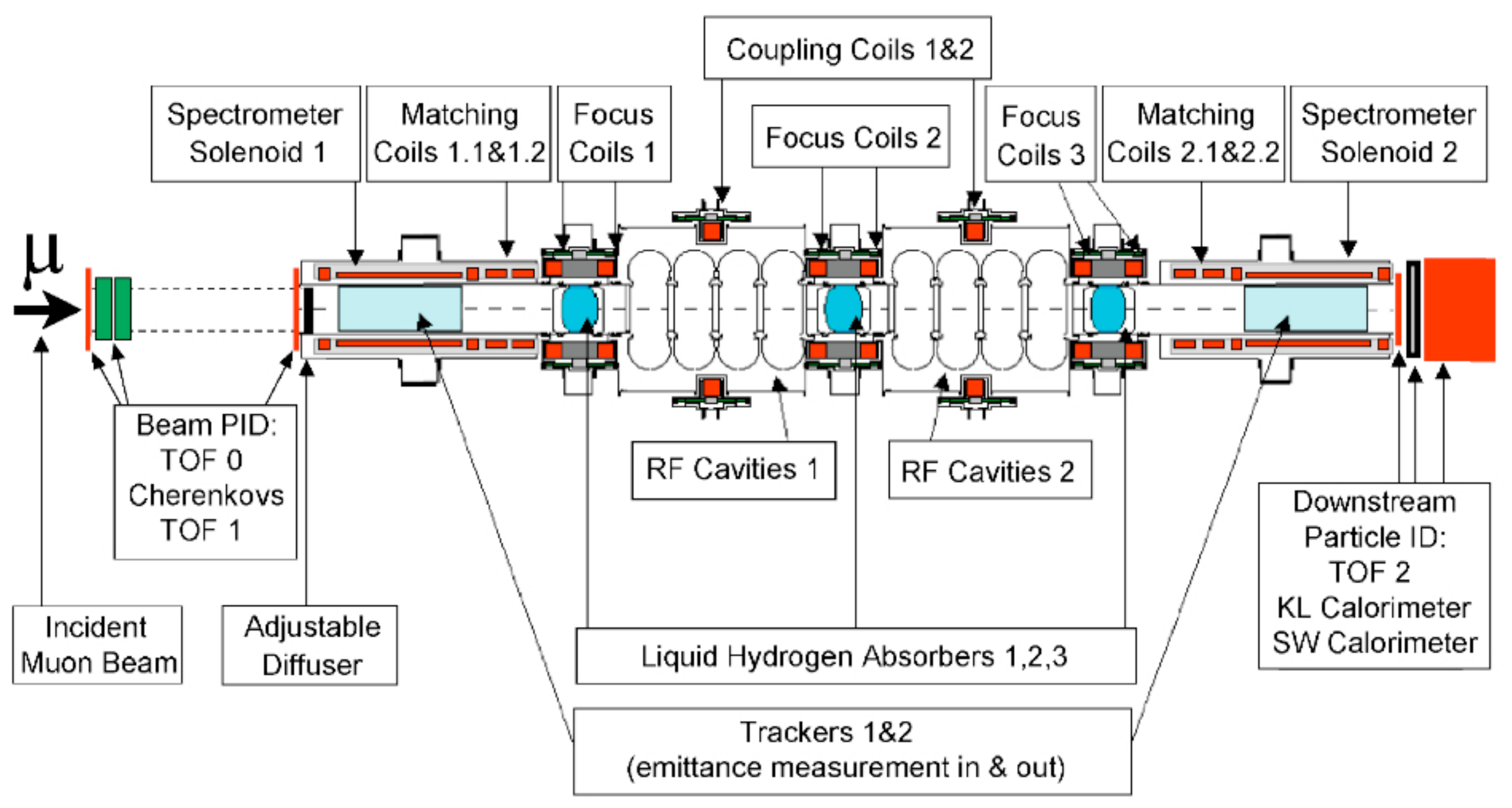}
\caption{Layout of the MICE experiment, with cooling cell in center surrounded by input and output spectrometers and particle-ID detectors. The muon beam is incident from the left.\label{fig:MICE}}
\end{figure}

The main components of MICE are shown in Fig.~\ref{fig:MICE}.  Cooling is provided by one cell of  the 2.75\,m cooling-channel  lattice from Neutrino Factory Feasibility Study-II~\cite{FS2}.\footnote{Some components of the Study-II cooling channel have been modified to reduce costs and to comply with RAL safety requirements.}  Incident muons first encounter two time-of-flight (TOF) counters, whose precise (50\,ps) time measurements contribute to particle identification (PID); together  with two aerogel threshold-Cherenkov counters, these reject background pions in the muon beam.  After this upstream PID, a variable-thickness lead diffuser generates a tunable input emittance.  

The muons next enter a spectrometer consisting of five stations of scintillating-fiber detectors within a uniform 4\,T solenoidal magnetic field; these measure the locations and momenta of each particle.  After this initial momentum measurement is the cooling section consisting of liquid-hydrogen absorbers, RF cavities, and superconducting coils.  An additional absorber at the end of the cooling section protects the downstream tracker from RF-cavity-generated x rays and dark currents.  Track positions and momenta after the cooling section are measured by a second spectrometer identical to the first.  Following the second spectrometer, a third TOF counter and two calorimeters provide further time and PID measurements to reject background electrons from muon decay.

\section{EMITTANCE}

The MICE cooling cell is designed to reduce the normalized transverse emittence of the muon beam by about 10\%, depending on the incident beam emittance and momentum and the magnetic configuration of the cell. The emittance of a beam is proportional to the phase-space volume occupied by the particles making up the beam. With the $z$ axis along the beam direction, the 6-dimensional normalized emittance is the square root of the 
determinant of the covariance matrix of the 6 phase-space coordinates $u_{6D} =(t, x, y, E, p_x, p_y)$ (scaled by the cube of the muon mass), and the 4-dimensional normalized emittance  is the square root of 
the determinant of the covariance matrix of transverse phase-space coordinates $u_{4D} =(x, y, p_x, p_y)$ (scaled by the square of the muon mass)~\cite{Emittance}.  The normalized transverse emittance $\epsilon_N$ is then the square root of the 4-dimensional normalized emittance. Defining the covariance matrices $(M_{ij})_{6D} = \sigma(u_i, u_j)_{6D}$ 
and $(M_{ij})_{4D} = \sigma(u_i, u_j)_{4D}$, 
we have
\begin{equation}
\epsilon_N^{6D}=\frac{
\sqrt{{\rm det}{(M_{ij})_{6D}}}}{(m_\mu c)^3}\,,\qquad
\epsilon_N^{4D}=\frac{
\sqrt{{\rm det}{(M_{ij})_{4D}}}}{(m_\mu c)^2}\equiv(\epsilon_N)^2\,.
\end{equation}
The rate of change of the normalized transverse emittance in a focusing magnetic lattice of beta function $\beta_\perp$ filled with an energy-absorbing medium of radiation length $X_0$ is given by~\cite{Neuffer}
\begin{equation}
\frac{d\epsilon_N}{dz}=-\frac{1}{\beta^2}\left|\frac{dE_\mu}{dz}\right|\frac{\epsilon_N}{E_\mu}+
\frac{\beta_\perp(0.014\,{\rm GeV})^2}{2\beta^3E_\mu m_\mu X_0}\,,
\end{equation}
where $E_\mu, m_\mu$ are  muon energy and mass, $\beta = v/c$, 
and $|{dE_\mu}/{dz}|$ is the mean rate of  muon energy loss in the medium.

\section{EMITTANCE MEASUREMENT}

The 10\% cooling provided by the MICE cooling cell is comparable to the emitance resolution of  standard beam instrumentation. MICE therefore requires precision (0.1\%) emittance measurements that can be achieved only via single-particle tracking.  MICE muon-momentum measurements are made with magnetic spectrometers.  At each fiber-tracking station, fiber-plane doublets in three views determine the $x$ and $y$ coordinates of each muon.  The muon momentum vector is reconstructed by fitting a helix to the measured positions. 

\begin{figure}
\vspace{-2.2in}
\includegraphics[width=.6\linewidth]{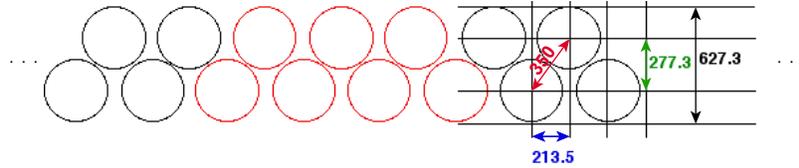}
\vspace{-2.4in}
\caption{Geometry of scintillating-fiber doublets (dimensions in $\mu$m); light from the seven center fibers (shown in red) is conveyed to a single VLPC pixel.\label{fig:ganging}}
\end{figure}

To minimize the contribution of multiple Coulomb scattering to the resolution, a scintillating-fiber diameter of only 350\,$\mu$m is used. With readout via cryogenic, solid-state, visible-light photon counters (VLPCs)~\cite{VLPC}, the mean number of photoelectrons observed is $\geq$10. To reduce the channel count, seven adjacent scintillating fibers are ganged to each VLPC pixel (Fig.~\ref{fig:ganging}). In order for the point resolution to be a small contribution to the emittance-measurement resolution, the r.m.s.\ resolution (430\,$\mu$m in each view) must not exceed about 10\% of the r.m.s.\ beam size, and similarly for angle measurement. 
To avoid biasing the emittance measurement, pions and decay electrons must be eliminated from the sample at better than the 0.1\% level; the MICE PID system is designed to achieve this. The TOF counters are also used to select muons for analysis that are near the crest of the 201\,MHz RF waveform.

The goal of MICE can be restated as vetting the simulation codes that will be used to design the actual muon-cooling channels for a neutrino factory or muon collider, by comparing the measured to the predicted cooling performance at each momentum and in each operating mode. In order to accomplish this, and to ensure measurement reproducibility and consistency, in addition to the precision particle-by-particle measurements just described, precise measurements of absorber and RF parameters (absorber thickness and density, RF phase and voltage) are also needed.

\begin{acknowledgments}
The authors thank their collaborators of the MICE Collaboration.
Work supported by National Science Foundation grant PHY-031737.
\end{acknowledgments}

\end{document}